\documentclass[aasms4,usenatbib]{mn2e}
\usepackage{epsfig}
\usepackage{graphicx}
\usepackage{amssymb}
\def\beq#1{\begin{equation}\label{#1}}
\def\eeq{\end{equation}}
\def\beqa#1{\begin{eqnarray}\label{#1}}
\def\eeqa{\end{eqnarray}}

\def\comment#1{\relax}

\begin{document}

\title[Settling accretion on to INS from ISM]
{Settling accretion on to isolated neutron stars from interstellar  
medium}

\author[S.B. Popov, K.A. Postnov, N.I. Shakura]
{ S.B. Popov $^{1}$, K.A. Postnov$^{1,2}$, N.I. Shakura$^{1}$
\thanks{E-mail: polar@sai.msu.ru (SBP)}\\
$^1${\sl Sternberg Astronomical Institute, Lomonosov Moscow State University,
Universitetskij pr. 13,
Moscow, 119991, Russia}\\
$^2${\sl Department of Physics, Lomonosov Moscow State University, Moscow,
Russia}
 }
\date{Accepted ......  Received ......; in original form ......
      }

\maketitle

\begin{abstract}

 We apply the model of subsonic settling accretion on to isolated
neutron stars accreting from the interstellar medium (AINS). 
We show that in this regime the expected mean X-ray luminosity from 
AINS turns out to be 
2-3 orders of magnitude as small as the maximum possible Bondi value,
i.e. $10^{27}-10^{28}$~erg s$^{-1}$. The intrinsically unstable character of
settling accretion due to long plasma cooling time leads to regular appearance of
X-ray flares with a duration of about one hour and a maximum luminosity of about the Bondi value, $\sim10^{31}$~erg~s$^{-1}$. This feature can be used to distinguished AINS from other dim X-ray sources. 
With the sensitivity of the forthcoming all-sky X-ray surveys
the expected number of the potentially detectable AINS can be from a few to ten.  

\end{abstract}

\begin{keywords}
stars: neutron --- accretion, accretions discs --- 
\end{keywords}

\section{Introduction}

 Already in the early 1970s it was proposed that old isolated neutron stars
(INSs) can reach the evolutionary stage at which their rotation is slow enough
to allow accretion from the interstellar medium (ISM) (\citealt{s71, ors70}). Despite early
optimistic estimates (\citealt{tc91}),
the ROSAT satellite failed to find any object from this
class (see a review of results of that time in \citealt{tr00}). 
This negative result has been thought either to be a consequence of very low
accretion luminosities due to large average
spatial velocities of NSs (\citealt{c98}), 
or due to a long time of spin evolution required for the NS to start accreting matter, 
which could be also because of large spatial velocities of INS (\citealt{census2000}). 
However, it was
demonstrated (\citealt{bp2010}) that INSs with initially large magnetic
fields, which are believed not to decay significantly 
below $\sim 10^{13}$~G, can become accreting isolated neutron stars (AINS),
and potentially can be detected by the eROSITA mission
(\citealt{2011SPIE.8145E..0DP}).   
 
Additional argument against the detectability of AINS by existing X-ray facilities
became popular when it was recognized that
the Bondi accretion rate $\dot M_\mathrm{B}$ is only an upper limit for the accretion rate
on to a INS moving in ISM. The actual
amount of matter that reaches the NS surface and  contributes to the
potential X-ray luminosity can be much lower than that derived from the 
Bondi rate. Such a reduction was found, for example, in numerical 
MHD simulations of accretion on to a magnetized INS (\citealt{toropina03}). 

Steady-state quasi-spherical accretion on to a gravitaitng center
was first considered by \cite{bondi1952}.
The exact solution of the hydrodynamical problem was obtained for adiabatic
gas motion. Both subsonic and supersonic accretion regimes were found.
Recently, the quasi-spherical accretion on to magnetized slowly rotating NS 
with account for cooling  and heating of the accreting gas was revisited in 
a series of papers by Shakura et al. (\citealt{2012MNRAS.420..216S,
shakura2013, spsp2014}, see 
\citealt{2014arXiv1407.3163S} for the latest review). 
It was recognized that at low X-ray luminosities
the captured matter, heated in the bow-shock, has no time to 
cool down and remains hot, which prevents it from entering the NS
magnetosphere via Rayleigh-Taylor (RT) instability (\citealt{el1977}).
Therefore, 
a hot and possibly convective quasi-spherical shell is formed around the magnetosphere.
The accretion rate through the shell at the stage of this settling subsonic accretion 
is determined by the ability of the plasma to enter the magnetopshere, 
which is controlled by the plasma cooling time: $\dot M_\mathrm{x}\simeq 
(t_\mathrm{ff}/t_\mathrm{cool})^{1/3}\dot M_\mathrm{B}$, where $t_\mathrm{ff}$ 
is the characteristic free-fall time
and $t_\mathrm{cool}$ is the plasma cooling time. The latter can be due to Compton processes (if the
photon field is sufficiently dense, e.g. at a high X-ray luminosity), or radiative processes, etc.
The regime of such a subsonic settling accretion can be realized in sources with X-ray luminosities
below $L_\mathrm{x}\lesssim 4\times 10^{36}$~erg s$^{-1}$. At higher luminosities, the matter 
cools down on a time scale shorter than the free-fall time and accretes 
supersonically towards the NS magnetosphere, and $\dot
M_\mathrm{x}=\dot M_\mathrm{B}$.   

Apparently, conditions for settling accretion are fulfilled for INSs accreting from ISM, 
and in this note we discuss the applicability of the settling accretion regime to this class of NSs. 
We show that the X-ray luminosity of such AINS  
is substantially reduced due to the lack of significant
cooling plasma mechanisms. When accretion from ISM starts, an AINS rapidly spins down due to 
the angular momentum transfer by viscous stresses in the surrounding shell, 
and therefore most of the AINS
are expected to be relatively dim long-period X-ray sources. 

\section{Settling accretion on to AINS}

We start with basic considerations related to accretion on to INSs from ISM.

 Despite low densities and high temperatures of the gas falling to   
the NS from ISM, accretion can be treated  
hydrodynamically (\citealt{sh1971}). Indeed, 
in the presence of magnetic field, the mean-free path of a proton is
determined by its Larmor radius:
$r_\mathrm{g}=m_\mathrm{p}v_\mathrm{T}c/(eB).
$
Here $m_\mathrm{p}$ and $e$ are the mass and the charge of a proton,
$v_\mathrm{T}$ is the component of the proton thermal velocity perpendicular
to the field line, 
$c$ is the speed of light, $B$ is the magnetic field. 
The relevant dimensionless number is the Knudsen number, which in our case can be determined  
as $\mathrm{Kn}=r_\mathrm{g}/R_\mathrm{A}$, where
$R_\mathrm{A}$ is the magnetosphere (Aflv\'en) radius.
As the thermal velocity of particles during accretion on to an INS cannot exceed the free-fall velocity,
$v_\mathrm{T}\le v_\mathrm{ff}(R_A)$, 
the Knudsen number can be re-written as the ratio of the Keplerian frequency
at $R_\mathrm{A}$ to the Larmor frequency:
$\mathrm{Kn}\lesssim \omega_\mathrm{K}/\omega_\mathrm{L}$.  
Numerically, $\omega_\mathrm{K}=\sqrt{GM/R_\mathrm{A}^3}\sim
10^{-2} \, \mathrm{rad} \, \mathrm{s}^{-1} (R_\mathrm{A}/10^{10}\, \mathrm{cm})^{-3/2}$ and
$\omega_\mathrm{L}=m_\mathrm{p}c/(eB)\sim 10^2 \, \mathrm{rad}\,
\mathrm{s}^{-1} B_\mathrm{\mu G}^{-1}$
(here $B_\mathrm{\mu G}$ is the magnetic field in microGauss). Therefore, 
for realistic conditions in an accretion flow on to a INS 
we have $\mathrm{Kn}\ll 1$, i.e. the mean-free path is much smaller than the scale of the problem, and
the hydrodynamic treatment of accretion is justified.

According to the classical Bondi-Hoyle-Littleton  accretion picture,
matter is gravitationally captured by a NS  
inside the Bondi radius:
\begin{equation}
R_\mathrm{B}=2GM/v^2\approx 4\times 10^{12} (M/M_\odot) v_7^{-2} {\rm cm}.
\end{equation}
Here $M$ is the NS mass,  
$v_7=v/10^7$~cm~s$^{-1}$ is the spatial velocity of the NS in ISM
(below we assume the NS velocity to be 
larger than the sound velocity in ISM, 
$v>c_\mathrm{s}\sim 10$~km s$^{-1}$, and so
the latter can be neglected).  

The corresponding Bondi-Hoyle-Littleton accretion rate is given by the standard relation:
\begin{equation}  
\dot M_\mathrm{B}=\rho_\infty v (\mathrm{\pi} R_\mathrm{G}^2)\sim 1.9\times 10^9
\, n \, v_7^{-3} \, {\mathrm g} \, {\mathrm s}^{-1},
\end{equation}
where $\rho_\infty$ is ISM matter density (typically 
$\rho_\infty=10^{-24}$~g~cm$^{-3}$, roughly corresponding to the 
particle number density $n\sim 1$~cm$^{-3}$).

In the settling accretion regime, the mass accretion rate on to the NS surface  
$\dot M_\mathrm{x}$ is controlled by the plasma cooling near the magnetopsheric boundary:
\begin{equation}  
\dot M_\mathrm{x}\simeq\dot M_\mathrm{B} \left(t_\mathrm{ff}/t_\mathrm{cool}\right)^{1/3}.
\end{equation}
Here the free-fall time $t_\mathrm{ff}$ is defined as $t_\mathrm{ff}=R^{3/2}/\sqrt{GM}$. 
The plasma cooling time inside the boundary layer above the NS magnetosphere, 
$t_\mathrm{cool}$, can be determined by different processes: 
Compton cooling, radiative losses, cyclotron emission, etc.
In the case of AINS we expect very low accretion rates and X-ray luminosities. 
Formally, in spherically symmetric photon fields 
the Compton cooling time becomes longer than the radiative cooling time only at very low X-ray luminosities, 
but as was explained in \cite{sph2013}, already at X-ray luminosities  
$\sim 10^{35}$~erg s$^{-1}$ and below, 
the X-rays generated near the NS surface should be 
narrowly beamed towards the magnetospheric cusp, which is more stable against 
the RT instability than the magnetospheric equator. Thus,  the accretion rate
will be 
determined by less efficient radiative plasma cooling near the magnetospheric equator. 

Near the magnetosphere, the cyclotron cooling can operate.
However, estimates show (see below) that the
cyclotron emission can also be neglected. At virial temperatures 
near the NS magnetospheric boundary $T\sim 10^7[\hbox{K}](R_\mathrm{A}/10^{10}[\hbox{cm}])^{-1}$
the plasma cooling is entirely due to free-free emission. 
In this case the cooling time is:
$$ 
t_\mathrm{cool}= 3\times 10^8 \left(\frac{R_\mathrm{A}}{10^{10}
\mathrm{cm}}\right) \times
$$
\begin{equation} 
                     \left(\frac{L_\mathrm{x}}{10^{30}
\mathrm{erg}\,\mathrm{s}^{-1}}\right)^{-1}
                     \frac{f(u)}{0.01} (1+X)^{-1} \, \mathrm{s}.
\end{equation}
Here $R_\mathrm{A}$ is the Alfv\'en radius, $L_\mathrm{x}$ is the X-ray
luminosity, $f(u)$ is the dimensionless factor determining the radial
settling velocity through the shell, $u_\mathrm{r}$, in units of the free-fall velocity, 
$f(u)=u_\mathrm{r}/u_\mathrm{ff}<1$ (essentially, the ratio $(t_\mathrm{ff}/t_\mathrm{cool})^{1/3}$).
%
%
By $X$ we denote possible additional contributions normalized to the bremsstrahlung losses.

The gas accreting from ISM necessarily brings chaotic magnetic fields,
therefore the cyclotron emission can contribute to the plasma cooling. 
The cyclotron losses can be estimated using equations from \cite{langer1982}.
It is easy to show that for reasonable assumptions about NS and ISM characteristics, the  
cyclotron losses are always significantly smaller than bremsstrahlung. 
Only in the case of INSs moving with very slow velocities ($v\lesssim10^6$~cm~s$^{-1}$), when 
$R_\mathrm{B}/R_\mathrm{A}\approx 10^4$, the value of $X$ can formally reach
unity for the fiducial ISM number density $n\sim 1$~cm$^{-3}$ and the magnetic
field in ISM $\sim$~few~$\mu$G.\footnote{Note that the 
reduction of $R_\mathrm{A}$ in INS with 
magnetic fields smaller than the standard $10^{12}$~G does not help much,
since INSs with small $\mu$ can hardly spin-down during the lifetime of the
Galaxy to start accreting from ISM (see \citealt{census2000, bp2010}).} 
Actually, large values of magnetic fields in the accreting plasma due to 
the magnetic flux
conservation in the accreting flow cannot be reached. In the first place, in
the settling envelope the magnetic field growth is expected to be slower than $r^{-2}$. In
addition, values $B>1$~G at the bottom of the envelope cannot be reached as
the magnetic field energy cannot be larger than the equipartition value ($B^2\sim 
v_\mathrm{ff}\dot M_\mathrm{B}/R_\mathrm{A}^2$). Therefore, 
below we will ignore any losses except free-free emission ($X\ll 1$).  

In the settling accretion model 
with radiative plasma cooling  \citep{sph2013} the Alfv\'en radius  is: 
%
\begin{equation}
R_\mathrm{A}\approx 2.2 \times 10^{10} L_{30}^{-2/9} \mu_{30}^{16/27}\,
\mathrm{cm}.
\end{equation}
Here $L_{30}=L_\mathrm{x}/10^{30}$~erg~s$^{-1}$.
The value of the dimensionless parameter
$f(u)$ is: 
\begin{equation}  
\label{furad}
f(u, \dot M_\mathrm{x})_\mathrm{rad}\simeq 0.005 \left(\dot
M_\mathrm{x}/10^{10}\,\mathrm{g}\,\mathrm{s}^{-1}\right)^{2/9} \mu_{30}^{2/27} (1+X)^{1/3}.
\end{equation}
In this expression $\mu=B_0R_{NS}^3/2$ is the NS dipole 
magnetic moment in units of
$10^{30}$~G~cm$^3$, $B_0$ is the NS polar magnetic field.
Thus, the expected accretion rate on to AINS is $\dot M_\mathrm{x}=f(u, \dot
M_\mathrm{x})_\mathrm{rad} \dot M_\mathrm{B}$.
Substituting here for $f(u)_\mathrm{rad}$ from Eq.(\ref{furad}) and
solving for $\dot M_\mathrm{x}$, we find:
\begin{equation}
\label{Mxrad}
\frac{\dot M_\mathrm{x}}{10^{10}\hbox{g\,s}^{-1}}\simeq 0.001 \mu_{30}^{2/21} (1+X)^{3/7} 
\left(\frac{\dot M_\mathrm{B}}{10^{10}\hbox{g\,s}^{-1}}\right)^{9/7}\,.
\end{equation}
This formula implies that the actual average accreting luminosity should be much 
smaller than the potential (Bondi) accretion luminosity of an AINS.

At this point a note on the character of accretion is in order. At low accretion rates, when the 
Compton cooling is ineffective, the average velocity of the magnetosphere plasma entry due
to RT instabily mediated by radiative cooling (see $f(u)_\mathrm{rad}$ above) is about 
one percent of the free-fall velocity. It is still much larger than the 
plasma entry velocities due to other possible mechanisms (diffusion, cusp instability and
chaotic magnetic reconnection, see e.g. \citealt{1984ApJ...278..326E}). Since the 
radiative cooling time in this case is $1/f(u)_\mathrm{rad}^3\sim 10^6$ times as 
long as the free-fall time $t_\mathrm{ff}(R_\mathrm{A})\sim 100$~s, the accretion should 
apparenlty proceed as a series of bursts with the duration determined by the RT time
instability, $t_\mathrm{inst}\sim t_\mathrm{ff}(R_\mathrm{A})/f(u)_\mathrm{rad}\sim 10^4$~s (see 
discussion in \citealt{shakura2013}). That is,
the system should ``wait'' for a fresh portion of hot accreting plasma 
to cool down enough for RT instability to fully develop. The peak luminosity during the outbursts
can be as high as during the Bondi accrtion, i.e. about $0.1 \dot
M_\mathrm{B} c^2\sim 10^{31}$~erg~s$^{-1}$
for typical parameters.


\section{Spin-down of AINS}

Now let us consider the spin-down of AINS in the settling accretion regime. 
Spin-down of AINS have been discussed in several papers (see, for example,
\citealt{ppk02} and references therein). The general conclusion is that AINS
should finally reach very long spin periods, mainly determined by
the properties of turbulence in ISM. 
In the model of settling accretion, these general conclusions do not
change qualitatively. However, quantitatively they are modified in the
sense that an AINS can spin-down faster due to more effective angular momentum
transport in the convective envelope surrounding the NS magnetosphere. 
Therefore, long spin
periods of AINS should be 
reached faster than, for example, in previous calculations by \cite{ppk02}.

To see the physics of the spin-down, 
we apply the basic equation describing viscous 
torques applied to 
a rotating ball of size $R$ immersed  
in a liquid with density $\rho$,
characterized by the dynamical viscosity coefficient $\rho v_\mathrm{l} l$
(see, for example, \citealt{ll1959}). Here  $v_\mathrm{l}$ is a
characteristic velocity and $l$ is a characteristic scale that determine the
viscosity. The ball spin evolution is then:  
\begin{equation}  
\frac{\mathrm{d}I\omega}{\mathrm{d}t}= -8 {\rm \pi} \rho
v_\mathrm{l}lR^3\omega,
\end{equation}
where $I$ is the moment of inertia of the ball and $\omega$ is the spin
frequency. 

If we substitute the ball by a gravitating body surrounded
by convective shell, the characteristic velocity should be 
$v_\mathrm{l}\sim v_\mathrm{ff}(R_\mathrm{A})$, and $l\sim R_\mathrm{A}$. 
Then we obtain:
\begin{equation}  
\frac{1}{\omega}\frac{\mathrm{d}\omega}{\mathrm{d}t}=-8 {\rm \pi}
\rho_\infty R_\mathrm{B}^{3/2}\sqrt{2GM}R_\mathrm{A}^2 I^{-1}.  
\label{sd}
\end{equation}
This formula corresponds to an exponential spin-down with 
the characteristic spin-down time:
\begin{equation}  
\tau_\mathrm{sd}=I \left[8 {\rm \pi}
\rho_\infty R_\mathrm{B}^{3/2}R_\mathrm{A}^2\sqrt{2GM}\right]^{-1}.
\end{equation}
Applying this formula to an AINS, we find that 
the spin-down time scale is:
$$
\tau_\mathrm{sd}=7.6 \times 10^5 I_{45} v_7^3 \left(\frac{M}{1.5 M_\odot}\right)^{-2}
\times
$$
\begin{equation}
\rho_{-24}^{-1} \mu_{30}^{-32/27} \left(\frac{L_\mathrm{x}}{10^{27}\,
\mathrm{erg}\, \mathrm{s}^{-1}}\right)^{4/9} \, \mathrm{yrs}. 
\end{equation}

According to Eq.(\ref{sd}) AINS can reach very long
(quasi) equilibrium periods (when the AINS spin is determined by the 
balance with turbulence in
ISM, convection in the envelope, etc.) much faster than
it was discussed by \cite{ppk02}. These authors numerically calculated the spin
evolution of a NS in ISM, and obtained that after starting accreting at
spin periods about hundreds of seconds, the NS reaches a critical period of
about few hours, when its behaviour starts to be
significantly influenced by ISM turbulence, in $\sim 6\times 10^7$~yrs.
Note also that \cite{ppk02} considered the model with linear decrease of the 
spin frequency, and in the situation we discuss here the spin-down is exponential.

\section{Discussion}

The above considerations suggest that 
the settling accretion from  ISM on to AINS naturally leads to 
low mean accretion luminosities $\dot M_\mathrm{x}\ll \dot M_\mathrm{B}$, in line with the lack of
AINS detection so far. It does not look promising to hope for their detection in the
near future as well, since the planned X-ray sensitivity, for example, of the 
SRG/eROSITA mission is expected to be $\sim 10^{-14}$~erg~s$^{-1}$~cm$^{-2}$
(\citealt{erosita}). Then  the expected X-ray luminosity of AINS
$L_\mathrm{x}\sim 10^{27}$~erg~s$^{-1}$ suggests a limiting distance of $30$~pc. 
According to \cite{bp2010}, only $\sim 10$ very dim candidates can be found 
in this volume, which are very difficult to identify.

However, the temporary enhancement of the accreting 
luminosity of AINS up to the Bondi value due to intrinsically intermittent
character of accretion when the plasma cooling time much exceeds the free-fall time, 
as described above, 
may help detecting such sources. Additionally,   
the accreting gas can bring large magnetic field loops. Once captured
into the accreting shell, the magnetic field will be enhanced close
to the base of the shell near the magnetosphere. If the field
in the loop is comparable with 
the magnetospheric field $B_\mathrm{m}\sim B_0(R_\mathrm{NS}/R_\mathrm{A})^3$,
the magnetic reconnection can happen. This would open the magnetospheric gates,
and the entire shell around the NS can be accreted in a free-fall
time from the Bondi radius. This mechanism was proposed by \cite{spsp2014} to explain the 
phenomenon of bright X-ray flares sporadically observed in an intensively 
studied subclass of high-mass X-ray binaries -- supergiant fast X-rat transients 
(SFXTs) accreting from stellar winds of massive optical early-type companion. 
Apparently, the similar scenario can be
applicable to the case of accretion from ISM on to INS. 

AINS are expected to have much lower steady X-ray luminosities 
($\lesssim 10^{30}$~erg s$^{-1}$, see Eq. \ref{Mxrad}) 
than SFXTs ($\sim 10^{34}$~erg s$^{-1}$) due to smaller settling radial
velocities of matter in the shell (i.e. smaller factors $f(u)$). The smaller 
$f(u)$, the longer the matter stays in the boundary layer near the magnetosphere
making it easier for reconnection to occur (see the discussion in \citealt{spsp2014}). 
Once the magnetospheric gates are opened, plasma entries the magnetosphere 
with the free-fall velocity, and accretion proceeds at about the Bondi
rate. Such a flare is expected to continue for about free-fall time from the Bondi
radius:
\begin{equation}
t_\mathrm{ff}=R_\mathrm{B}/v(R_\mathrm{B})=2GM/v^3\approx 4\times 10^5 
v_7^{-3} \, \mathrm{s}.
\end{equation} 
Note that the X-ray luminosity during such a flare 
can be variable on the time scale $\sim
R_\mathrm{A}/v(R_\mathrm{A})\sim 100 \, {\mathrm s}$ for realistic
conditions when $R_\mathrm{A}\sim 10^{10}$~cm.  

Therefore, during X-ray flares lasting for about from several hours to one day
(when the entire shell is unstable) the X-ray luminosity 
from AINS can reach $L_\mathrm{x}\sim 0.1 \dot M_\mathrm{B}c^2 \sim 10^{31}$~erg s$^{-1}$. 
From a fiducial distance of 100~pc, this would correspond to an X-ray flux of 
$\sim 10^{-11}$~erg~cm$^{-2}$~s$^{-1}$, well within the reach of the modern 
X-ray telescopes. It would be intriguing to search for such X-ray flares in the existing
X-ray archives.

It is non-trivial to estimate the number of the potentially observable flares. We expect
(see also \citealt{popov2000b}) that the spectrum from AINS would be thermal with
typical temperatures $\lesssim 1$~keV. For such soft sources the operating Swift/BAT all-sky monitor is
not very effective. The RXTE All-Sky Monitor (AMS) had a limiting sensitivity of $\sim 20$~mCrab, which
corresponds to a minimum detectable flux of $\sim 4 \times 10^{-10}$~erg~cm$^{-2}$~s$^{-1}$. If we
neglect interstellar absorption (which may be well justified for small 
distances around 100 pc), then the limiting distance to AINS reduces to $\sim 15$~pc. 
With a local
density of AINS of $\sim 10^{-4}$~pc$^{-3}$ (\citealt{bp2010}) only a few sources
can be found within this volume. The rate of possible flares short flares (with
a duration of $t_\mathrm{inst}\sim $~a few hours) can be rather high, but the rate of longer bursts
caused by magnetic reconnection is difficult to estimate.  The MAXI monitor onboard the
International Space Station has observed the sky in the range 0.5-30 keV for 5
years. 
Potentially, its sensitivity for a one-day observation could be about a few
mCrabs, however, detectors were gradually degrading, and the dimmest X-ray
nova detected by this instrument have a peak flux of  $\sim100$~mCrab (see, for example,
\citealt{mihara}). With such a high flux threshold, again, we are left with a very small
number of potentially observable sources. Additionally, the local ISM (inside
$\sim 100$~pc) is dominated by the Local Bubble (see e.g. \citealt{posselt} for the 
discussion of a 3D model of the local ISM) filled with a  hot low-density gas, and
therefore the appearance of accreting isolated NSs can be greatly suppressed inside this volume. 
Probably, AINS (flaring or quasi-stationary) could be identified as serendipitous
sources in the deep Chandra and XMM-Newton exposures, but the analysis of this
possibility is beyond the scope of this note.

\section{Conclusions}

In this short communication we discuss the observational signatures of 
isolated accreting NSs in the
model of settling subsonic accretion, which is expected to naturally occur on to 
slowly rotating magnetized NS in ISM. It is shown that the average 
accretion luminosity of AINS should be
significantly reduced relative to the Bondi accretion luminosity. 
Such NSs are expected to reach very long spin-periods in relatively
short time $\lesssim 10^6$~yrs, after which their spin behaviour is mainly
determined by the interstellar turbulence. 

AINS can potentially be detected as transient X-ray sources due to the  
the intrinsically unstable character of settling accretion from magnetized ISM.
The X-ray flares can last from  hours up to about one day with a peak
luminosity of $\sim 10^{31}$~erg~s$^{-1}$, 
and can be searched for in the forthcoming X-ray survey mission like eROSITA. 
We stress that the presence of the characteristic time of order of one
hour in the temporal X-ray variability can be a distinctive feature of AINS, which 
can be used to separate them from stellar flares that must show different time properties.

\section*{Acknowledgements} 

The work of SBP and NIS is supported by the Russian Science Foundation grant 14-12-00146.
KAP also acknowledges the support from RFBR grant 14-02-00657.

\bibliography{bib_ains2014}

\end{document}